\documentclass[aps,prx,twocolumn]{revtex4-2}

\usepackage{epsfig}
\usepackage{amssymb}
\usepackage{graphicx}

\begin{document}

\title{Gap resonance in the classical dynamics of the current-biased Josephson tunnel junctions}
\author{Dmitri V. Averin}
\affiliation{Department of Physics and Astronomy, Stony Brook University, SUNY,
Stony Brook, NY 11794-3800}

\begin{abstract}

This work reports a novel time-domain expression for the current-response kernels of a Josephson tunnel junction between BCS superconductors with in general different energy gaps, and its use to simulate the classical dynamics of such junctions. The simulations show a dynamic regime characterized by the resonance between the Josephson oscillations and the gap oscillations in the asymptotics of the current kernels that has not been studied previously. The resonance manifests itself as a hysteresis in the dc current-voltage characteristics of the current-biased junctions in the voltage range above the energy gap, in addition to the usual ``inertial'' hysteresis characteristic for tunnel junctions at voltages below the energy gap. Features of the IV curves related to the gap resonance, including the above-the-gap hysteresis, should manifest themselves in many structures and devices utilizing high-quality Josephson tunnel junctions with relatively large, but achievable current densities.

\end{abstract}

\maketitle

\section{Introduction}

\label{sec1}

A tunnel junction: two conductors separated by a layer of insulator that is sufficiently thin to allow for charge transfer between the conductors at a rate that is large enough to produce a noticeable current \cite{giaever,cohen}, has a long and still continuing history as a source of novel phenomena in physics, and a tool for both practical devices and scientific experiments. Probably the best studied and widely used are the Josephson tunnel junctions between superconductors, which exhibit Josephson effect \cite{josephson,anderson}, a quantum-coherent transfer of Cooper pairs, which leads to supercurrent flow across the tunnel barrier. An important feature of this effect, which makes possible its various applications, is that the Josephson supercurrent has the same magnitude as the typical current associated with electron tunneling in the junction \cite{nicol,josephson}. This is possible because the magnitude of the supercurrent, as a quantum-coherent process, is determined by the amplitude of tunneling, in contrast to the probability of tunneling for electron current. The amplitude of the two-electron transfer of a Cooper pair has the same magnitude as the probability $D$ of one-electron transmission through a tunnel barrier, and is very small, $D\ll 1$, for typical tunnel barriers.

Josephson tunnel junctions possess several other characteristics, starting with the strongly nonlinear dynamics of the dynamic junction variables: Josephson phase difference $\phi(t)$ and the voltage $V(t)$ across the junction, that make them ideal building blocks for various devices of superconductor electronics, including digital circuits \cite{likharev1991,tolpygo}. Another features of the high-quality tunnel junctions with $D\ll 1$, is that there is a wide range of voltages $V$ around zero, where the dissipative electron tunneling is strongly or, ideally, completely suppressed, and the junction dynamics is reduced to the quantum-coherent Cooper-pair tunneling. Tunnel junctions in this regime demonstrate various effects of the macroscopic quantum dynamics of the Josephson phase $\phi$ \cite{leggett,azl,cp1,friedman,cp2,cp3,cp4}, which at present play most important role in the development of superconducting quantum computing circuits (see, e.g., \cite{clarke,dwave,devoret,martinis1,martinis2}). As a less daunting task, the same suppression of the dissipative electron tunneling makes it possible to use Josephson tunnel junctions for the development of classical circuits for thermodynamically reversible computing \cite{rev}.

Classical current response of a Josephson tunnel junction to an arbitrary fixed time-dependent voltage $V(t)$ is well understood for a long time \cite{werthamer,larkin} and is described in details in classic textbooks \cite{barone,likharev}. Nevertheless, although the assumption of the fixed bias voltage $V(t)$ is natural for the  theoretical calculation of the current in the junction, in practice, typical bias conditions are quite different, with the simplest case being the current bias. Nonlinear nature of the Josephson dynamics makes the task of analyzing the junction transport properties in these realistic situations quite nontrivial even if the basic voltage-biased current response is known. Since the general nonlinear dynamics of the Josephson phase $\phi$ is described more naturally in time domain, additional layer of difficulty is added to this problem by the fact that in the basic theoretical description  \cite{werthamer,larkin}, the current response is more naturally calculated in the energy (i.e., frequency) representation. Expression for the current response of a Josephson tunnel junction directly in time domain was obtained only for junctions with equal energy gaps of the two electrodes \cite{harris1976}. This work suggests a more general expression for the time-domain current response of a Josephson tunnel junction between two BCS superconductors with different energy gaps. This expression makes it possible to analyze more conveniently Josephson dynamics of the current-biased junctions and calculate the current-voltage characteristics of these junctions. The calculations show a novel ``gap resonance'' regime of the Josephson dynamics in current-biased tunnel junctions.

Next Section discusses the main result of the first part of this work, expression for the time-domain current response of a tunnel junction, which is derived in Appendix A. Section III describes the application of this expression to the simulation of the Josephson dynamics of the current-biased tunnel junctions and calculates the current-voltage characteristics of the junction under several different conditions. Section IV estimates the junction parameters under which the gap-resonance regime of the Josephson dynamics is realized, and concludes. Appendix B gives some details of the numerical procedure used in the simulations of the junction dynamics.

\section{Current kernels}

Consider a Josephson tunnel junction between the two BCS superconductors with in general different energy gaps $\Delta_1$ and $\Delta_2$. If the bias voltage $V(t)$ between the superconductors has an arbitrary dependence on time $t$, the tunnel current $I_T(t)$ that flows in the junction can be expressed through the time-dependent Josephson phase difference $\phi(t)$ across the junction (see, e.g., \cite{barone,likharev}):
\begin{eqnarray}
I_T(t)=\int_{-\infty}^t dt' \Big[I_p(t-t')\sin\frac{\phi(t)+\phi(t')}{2} \;\;\;\;\;\;\;\; \nonumber \\  +I_{qp}(t-t') \sin\frac{\phi(t)-\phi(t')}{2}\Big] \, ,\;\;\;\; \dot{\phi}(t) = 2eV(t)/\hbar \, . \;\;\; \label{eq1} \end{eqnarray}
As shown in Appendix A, the pair current $I_p$ and the quasiparticle $I_{qp}$ kernels in this equation have the following analytical form for the tunnel junction with the normal-state resistance $R_N$:
\begin{eqnarray}
I_p(t)=-\frac{\pi \Delta_1 \Delta_2}{2e\hbar R_N} \Big[ J_0\big(\frac{\Delta_1 t}{\hbar}\big) A_0\big(\frac{\Delta_2 t}{\hbar},\beta_2\big)+A_0\big(\frac{\Delta_1 t}{\hbar},\beta_1 \big) \nonumber \\ \cdot J_0 \big( \frac{\Delta_2 t}{\hbar}\big)\Big]\, ,\;\;\; I_{qp}(t)=\frac{-\hbar}{eR_N}\delta'(t-0)+ \frac{\pi \Delta_1 \Delta_2}{2e\hbar R_N}\;\;\;\;\; \label{eq2}  \\ \cdot \Big[ J_1\big(\frac{\Delta_1 t}{\hbar}\big) A_1\big(\frac{\Delta_2t}{ \hbar},\beta_2\big) + A_1\big( \frac{\Delta_1 t}{\hbar},\beta_1\big) J_1 \big(\frac{\Delta_2 t}{\hbar}\big)\Big] \, . \;\;\; \nonumber \end{eqnarray}
The functions $J_{0,1}$ here are the Bessel functions of the first kind, which for $z>0$ can be defined by the following relations \cite{as}:
\begin{equation}
J_0(z)=\frac{2}{\pi} \int_1^{\infty} \frac{dx \sin zx}{\sqrt{x^2-1}}\, , \;\;
J_1(z)=-\frac{d}{dz} J_0(z) \, ,
\label{bes} \end{equation}
while the functions $A_{0,1}$ depend on the temperature $T$ of the superconductors through the ratios $\beta_j\equiv \Delta_j/2T$, $j=1,2$, and are defined, also for $z>0$, as
\begin{equation}
A_0(z,\beta)=-\frac{2}{\pi} \int_1^{\infty} \frac{dx \cos zx}{\sqrt{x^2-1}}\tanh \beta x \, , \label{eq3} \end{equation}
\[ A_1(z,\beta)=-\frac{d}{dz}A_0(z,\beta)= -\frac{2}{\pi} \int_1^{\infty} \frac{dx x\sin zx}{\sqrt{x^2-1}}\tanh \beta x \, . \]
The first, $\delta'$ term in the quasiparticle kernel $I_{qp}$ (\ref{eq2}) accounts for the normal-state Ohmic contribution $I_N$ to the current: $I_N(t)= V(t)/R_N$.

At low temperatures, when $\tanh \beta x \equiv 1$, the functions $A_j$ are reduced to the Bessel functions $Y_j(z)$, $j=0,1$, of the second kind:
\[ A_j(z,\beta\rightarrow \infty)=Y_j(z)\, .\]
As a result, in the low-temperature limit, the current kernels (\ref{eq2}) are expressed through the well-studied Bessel functions only:
\begin{eqnarray}
I_p(t)=- \frac{\pi \Delta_1 \Delta_2}{2e\hbar R_N} \Big[ J_0\big(\frac{\Delta_1 t}{\hbar}\big) Y_0\big(\frac{\Delta_2 t}{\hbar}\big)+Y_0\big(\frac{\Delta_1 t}{\hbar}\big) J_0 \big( \frac{\Delta_2 t}{\hbar}\big)\Big] ,\nonumber \\  I_{qp}(t)=\frac{-\hbar}{eR_N}\delta'(t-0)
\;\;\;\;\;\;\;\;\;\;\;\;\;\;\;\;\;\;\;\;\;\;\;\;\;\;\;\;\; \label{eq5} \end{eqnarray}
\vspace{-3ex}
\[ + \frac{\pi \Delta_1 \Delta_2}{2e\hbar R_N} \Big[ J_1\big(\frac{\Delta_1 t}{\hbar}\big) Y_1\big(\frac{\Delta_2t}{\hbar}\big) + Y_1\big( \frac{\Delta_1 t}{\hbar}\big) J_1 \big(\frac{\Delta_2 t}{\hbar}\big)\Big] \, . \]
This result generalizes to different energy gaps $\Delta_1\neq \Delta_2$ the equations obtained by Harris \cite{harris1976} for equal gaps. Also, one can check that the pair and the quasiparticle kernels (\ref{eq5}) reproduce the frequency-domain expressions for these kernels at $T=0$ in terms of the elliptic integrals \cite{werthamer,larkin} in the cases for which the corresponding integrals of the Bessel functions \cite{prudnikov} are available (see, e.g., Eq.~(\ref{aic}) below).

As a brief aside, one can note that the generalization to different energy gaps implies that Eqs.~(\ref{eq1}) to (\ref{eq5}) can be used to describe the time-dependent current response of a normal metal/superconductor (NS) junction in the tunnel limit. Indeed, for $\Delta_2=0$, the pair current $I_p$ vanishes, while in the limit $\Delta_2\rightarrow 0$, the quasiparticle kernel is obtained from $A_1 (z,\beta_2)$ evaluated for $z,\beta_2\rightarrow 0$, with $z/\beta_2=2Tt/\hbar$. For $z\rightarrow 0$ (but $z\neq 0$), the integral $A_1 (z,\beta_2)$ is determined by the range of large values of the integration variable, where it is reduced to
\begin{eqnarray}
\Delta_2 A_1(z,\beta_2)=-\frac{2\Delta_2}{\pi} \int_0^{\infty} dx \sin zx \tanh \beta_2 x e^{-\epsilon x}\Big|_{\epsilon \rightarrow 0} \nonumber\\
=-\frac{\Delta_2}{\pi \beta_2} \sum_{n=-\infty}^{\infty}\frac{(-1)^n z/2\beta_2}{n^2+ (z/2\beta_2)^2} =-\frac{2T}{\sinh(\pi Tt/\hbar)}  \, . \;\;\; \label{eq6} \end{eqnarray}
This gives for the current-response kernel of the NS tunnel junction:
\begin{equation}
I_{qp}^{(NS)}(t) = \frac{-\hbar}{eR_N}\delta'(t-0) - \frac{\pi T \Delta}{e\hbar R_N}
\frac{J_1(t\Delta/\hbar)}{ \sinh (\pi Tt/\hbar)} \, .
\label{tun} \end{equation}
This expression coincides with the $D\rightarrow 0$ limit of the general expression for the time-dependent response of an NS junction with an arbitrary electron transparency $D$ \cite{averin2020}.

The next Section describes the application of the expressions for the current kernels presented in this Section to the simulations of the classical Josephson dynamics of the current-biased junctions. An outline of the derivation of these expressions is discussed in Appendix A.

\section{Gap resonance in the dynamics of the current-biased junctions}

\label{sec3}

As an application of the current kernels obtained in the previous Section, I now discuss the simulations of the classical dynamics of a Josephson phase $\phi$ across a current-biased Josephson tunnel junction formed by the BCS superconductors with in general different gap energies $\Delta_1$ and $\Delta_2$. In this work, the simulations are limited to the low-temperature regime, when the current kernels are given by Eq.~(\ref{eq5}) containing only the Bessel functions. The ``brute-force'' approach to the simulation of the junction dynamics employed here consists of the direct numerical evaluation of the current kernels (as described in the Appendix B), their direct numerical integration at each step of the time evolution, and subsequent solution of the integro-differential equations for the junction dynamics that are described below. This approach differs from the existing approaches to the simulation of the Josephson dynamics in the tunnel-junction model \cite{odintsov1987,gulevich2017,pscan2,gulevich2020}, which use the scheme that avoids the need to solve the integro-differential equations, and in particular, to evaluate the integrals (\ref{eq1}) for the tunnel current at each time step of the Josephson evolution. The scheme is based on the approximate representation of the current kernels by the sums of the exponentials, which effectively makes it possible to take the integrals analytically. An important advantage of this computation scheme is that it is very efficient and enables one to treat practical present-day circuits of superconductor electronics with the number of junctions that can reach into hundreds of thousands. On the other hand, because of the singular nature of the current kernels which are characterized by the very slow and oscillatory decay with time, this approach can create some uncertainty as to whether the features observed in the numerical simulations are real or the consequences of the approximations used for the current kernels \cite{kirichenko}. While the brute-force scheme adopted in this work is much slower, it does not create this uncertainty. The main new result of the simulations presented below is the demonstration of the ``gap resonance'' regime of the junction dynamics which produces the hysteresis in the current-voltage characteristics of the current-biased junction appearing in the range where the dc component of the voltage $V$ across the junction is above the gap value $V_g=(\Delta_1+\Delta_2)/e$. In what follows, I present the results of the detailed numerical simulations of the dynamics of the current-biased Josephson tunnel junctions that show the gap resonance, and provide the semi-quantitative analytical explanation of the above-the-gap hysteresis.

\begin{figure}[t]
\centering
\includegraphics[width=0.49\textwidth]{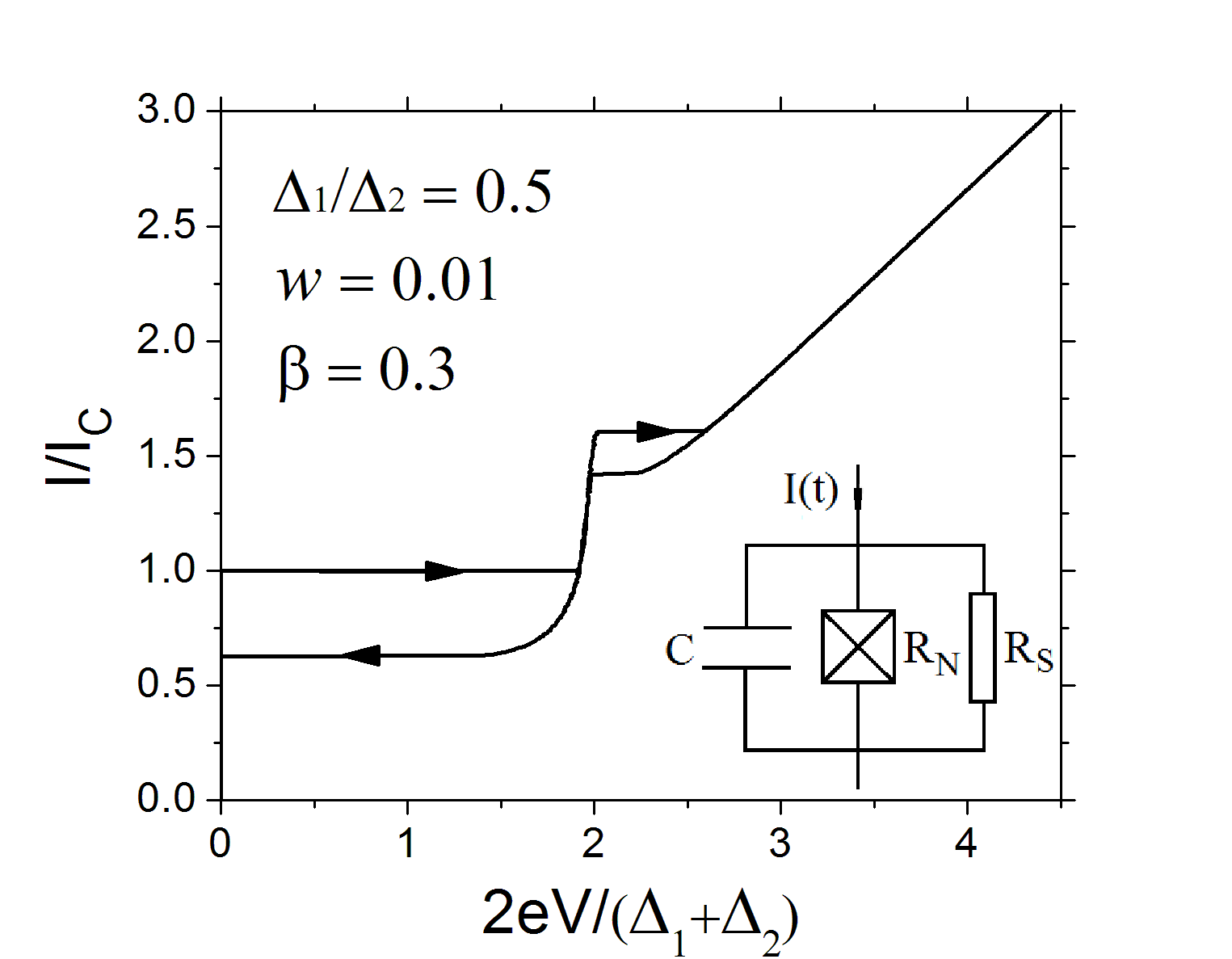}
\caption{{\protect\footnotesize {Current-voltage characteristic of a current-biased Josephson tunnel junction between two BCS superconductors. Parameters $\beta$ and $w$ are defined by Eqs.~(\ref{bet}) and (\ref{gbr}), respectively. } }}
\label{fig1} \end{figure}

Classical Josephson dynamics of a Josephson tunnel junction shunted by an Ohmic resistor $R_S$ and biased by in general time-dependent external current $I(t)$ (see the equivalent circuit in the inset in Fig.~\ref{fig1}) is governed by the standard set of coupled dynamic equations for the phase $\phi$ across the junction and the instantaneous voltage $V$ on the junction capacitance $C$:
\begin{equation}
\dot{\phi}=\frac{2eV}{\hbar }\, ,\;\;\; C\dot{V}=I(t)-\frac{V}{R_S}-I_T(t) \, .
\label{dyn} \end{equation}
Here, the tunnel current $I_T$ is given by Eq.~(\ref{eq1}) with the current kernels (\ref{eq2}). Since the calculation of tunnel current $I_T$ (\ref{eq1}) requires integration over time, the resulting  evolution equations (\ref{dyn}) for the phase $\phi$ are the integro-differential equations. The first step towards their numerical solution is to express them in the dimensionless form. To do this, it is natural to normalize time $t$ to the frequency $\Omega$ of the gap oscillations, $\Omega=(\Delta_1+\Delta_2)/\hbar$, and the voltage $V$ -- to half of the gap voltage $V_t=(\Delta_1+\Delta_2)/e$. Then, it is convenient to introduce the auxiliary dimensionless parameters $\alpha =1+R_N/R_S$ characterising the degree to which the junction is shunted, with $\alpha=1$ describing the unshunted junction, and $\lambda = (\Delta_1+\Delta_2)^2/ (\pi\Delta_1 \Delta_2 )$ characterising the magnitude of the difference between the two energy gaps of the junction electrodes. With this definition, $\lambda = 4/\pi$ corresponds to the junction with equal energy gaps.

The main physical parameter that controls the nature of the phase dynamics in the tunnel junction is the ratio of the time constant $R_NC$ of the junction capacitance to the gap time:
\begin{equation}
\beta=\frac{\Delta_1+\Delta_2}{\hbar}R_N C\, .\label{bet}
\end{equation}
For the purpose of the discussion of the junction dynamics below, it will be convenient to view $\beta$ simply as the parameter characterizing the magnitude of the junction capacitance, ranging from the negligibly small capacitance at $\beta=0$ to large capacitance for $\beta\gg 1$. It should be noted, however, that this view is misleading in one respect. For typical tunnel junctions, the time constant $R_NC$ (and therefore $\beta$) is independent of the junction area, in contrast to the absolute magnitude of the junction capacitance, because the area dependence of the junction resistance $R_N$ compensates that of the capacitance. Both the time constant $R_NC$ and $\beta$ depend only on the effective electron transparency of the tunnel barrier that is characterized usually through the ``critical current density'', decreasing with increasing current density as the inverse of it.

With this set of the dimensionless parameters and variables, the dynamic equations (\ref{dyn}) with the appropriate current kernels take the form:
\begin{eqnarray}
\beta \lambda \dot{v}=i(t)-\alpha \lambda v +\int_0^{\infty} d\tau \Big[ i_p(\tau)\sin \frac{\phi(t)+\phi(t-\tau)}{2} \nonumber \\
 - i_{qp} (\tau)\sin \frac{\phi(t)-\phi(t-\tau)}{2}\Big] \, ,\;\;\;\;\;\; \dot{\phi}=v\, ,
\;\;\;\;\;\;\; \label{dim} \end{eqnarray}
where both times $t$ and $\tau$ are normalized to the gap frequency, and the current kernels are expressed through the Bessel functions:
\begin{eqnarray}
i_p(\tau)=J_0(r_1 \tau)Y_0(r_2 \tau) +J_0(r_2 \tau)Y_0(r_1 \tau)\, , \nonumber \\
i_{qp}(\tau)=J_1(r_1 \tau)Y_1(r_2 \tau) +J_1(r_2 \tau)Y_1(r_1 \tau)\, . \label{dim2}
\end{eqnarray}
Here $r_j$ are the partial gaps: $r_j=\Delta_j/(\Delta_1+\Delta_2)$, $j=1,2$, i.e., $r_1+r_2=1$.

Conventions adopted above dictate that the current in the junction is normalized to
\begin{equation}
I_N=\frac{\pi \Delta_1\Delta_2}{2e(\Delta_1+\Delta_2)R_N}\, ,\label{ain} \end{equation}
e.g., $i=I/I_N$. The normalization current $I_N$ is different from the junction critical current $I_C$, which is expressed in terms of the complete elliptic integral of the first kind $K(m)$ as (see, e.g., \cite{barone,likharev})
\begin{equation}
I_C=\int_0^{\infty} dtI_p(t)=\frac{2 \Delta_1\Delta_2}{e(\Delta_1+\Delta_2)R_N}K \Big(\frac{|\Delta_1-\Delta_2|}{\Delta_1+\Delta_2}\Big) \, .\label{aic} \end{equation}
For completeness, one should note that the elliptic integral in Eq.~(\ref{aic}) is defined as in Ref.~\cite{prudnikov}, not as in Ref.~\cite{as}, in which case the argument of $K$ should be squared. In the case of equal gaps, the relation between the two currents simplifies to $I_N=I_C/2$.

One of the main qualitative features of the current in a Josephson tunnel junction are the singularities (peaks and jumps) of the current components resulting from the singularity of the density of state in the superconducting electrodes at the edge of the energy gap. In the time domain, this gap-edge singularity manifests itself in a very slow decay of the quasiparticle and the pair current kernels. Using the standard expressions for the leading asymptotic behavior of the Bessel functions at large arguments, we see that at large time, $\tau\gg 1$, the current kernels (\ref{dim2}) simplify to:
\begin{equation}
-i_p(\tau)=i_{qp}(\tau)=\frac{2}{\pi \sqrt{r_1r_2}} \frac{\cos \tau}{\tau}\, .
\label{lt} \end{equation}
Although this expression is derived for the current kernels at vanishing temperature, the finite temperature $T$ does not change the slow-decaying nature of this asymptotic behavior. Using Eq.~(\ref{eq3}) for the finite-temperature regime, one can show that the amplitude of the asymptotic (\ref{lt}) decreases as $\tanh \Delta/T$ as a function of temperature $T$, while its time dependence remain the same. This means that the problems of divergence resulting from the slow decay of the current kernels are not cured by finite temperatures. This conclusion is qualitatively consistent with the fact that the long-time asymptotic behavior (\ref{lt}) of the current kernels is produced by the singularity of the superconducting density of states at energy $\Delta$ which is not affected by the temperature. Not too close to the superconducting transition, the temperature changes only the occupation of the states at the gap edge, which leads to the decrease of the overall magnitude of the asymptotics (\ref{lt}) without affecting its dependence on time.

Expression (\ref{lt}) for the long-time asymptotics of the current kernels shows that the main integral (\ref{eq1}) for the tunnel current is not absolutely convergent. When the evolution of the Josephson phase is such that the function $\cos \tau$ in the asymptotics (\ref{eq1}) taken together with the corresponding functions of the Josephson phase in Eq.~(\ref{eq1}) has a non-vanishing average over the period of the gap oscillations, the tunnel current diverges logarithmically because of the slow, $1/\tau$, decay of the current kernels (\ref{lt}) with time. At fixed bias voltage, this leads to the logarithmic ``Riedel'' singularity \cite{riedel} of the supercurrent amplitude at the gap voltage. Of course, in real Josephson junctions, this singularity is smeared by several mechanisms (see, e.g., \cite{zorin1979}), the two most basic one being small static variations of the energy gap over the area of the junction, and electron-phonon relaxation producing finite lifetime of the quasiparticle energy states in the superconducting electrodes of the junction. In junctions with finite electron transparency $D$ (assumed to be vanishingly small in tunnel junctions), finite $D$ itself directly smears the Riedel singularity \cite{averin1995}. Smearing of any kind turns Riedel singularity into a finite ``Riedel peak'' of the supercurrent, which was actively studied since the early days of investigations of the Josephson effect -- see, e.g.,  \cite{hamilton1971,buckner1972,kofoed1975,vernet1976,morita1983,winkler1993}. These investigations showed that the typical magnitude of the energy gap smearing in tunnel junctions is quite small, on the order of 0.01 of the energy gap itself. Different gap smearing mechanisms produce different specific time dependence of the decoherence factors for the gap oscillations. For instance, static variations of the gap over the junction area results in the Gaussian decay of the gap oscillations, while the finite lifetime of the quasiparticle states, in the most basic approximation, leads to the exponential decay. For realistic weak smearing of the gap, however, precise nature of the smearing is not important. In time domain, weak smearing affects only the long-time asymptotic behavior (\ref{lt}) of the current kernels. It suppresses the logarithmic divergence of the current by effectively limiting the integration range in Eq.~(\ref{eq1}) for the tunnel current to some finite time $t_0$. As a result, only this cut-off time, and not the precise time profile of the cut-off is important in the regime of weak gap smearing, when $t_0$ is much larger than the gap oscillation period $2\pi/\Omega$.

This means that in order to obtain a well-defined tunnel current $I_T(t)$, and from this, a consistent description of the Josephson dynamics of a tunnel junction based on Eqs.~(\ref{dyn}), (\ref{dim}, and (\ref{dim2}), one needs to include in the model a certain smearing mechanism of the energy gap which, in time domain, ensures a faster decay of the current kernels than in the ideal BCS regime described by Eq.~(\ref{lt}). Numerical simulations discussed below assume a model of small Gaussian fluctuations of the gap across the junction area, which produces an extra Gaussian factor multiplying the current kernels (\ref{eq2}) and improving convergence of the current integrals (\ref{eq1}). When the kernels are expressed through the dimensionless time $\tau$, as in Eq.~(\ref{dim2}), the pair kernel is transformed then like this:
\begin{equation}
i_p(\tau)\;\;\; \rightarrow \;\;\; e^{-w^2\tau^2}i_p(\tau),
\label{gbr} \end{equation}
with the similar transformation of the quasiparticle current kernel  $i_{qp}(\tau)$. Parameter $w$ in Eq.~(\ref{gbr}) characterizes the gap smearing, and effectively cuts off the current kernels at time $t_0\simeq 1 /(\Omega w)$. As follows from the experimental results \cite{hamilton1971,buckner1972,kofoed1975,vernet1976,morita1983,winkler1993}, one can use an estimate $w\simeq 0.01$, although the gap smearing strength varies for different junction materials and structures. Such a Gaussian suppression factor for the gap oscillations in time-domain implies that the magnitude of the total energy gap of the junction exhibits small Gaussian fluctuations around the mean value $\Delta_1+\Delta_2$ of relative magnitude $w$.

\begin{figure}[t]
\centering
\includegraphics[width=0.49\textwidth]{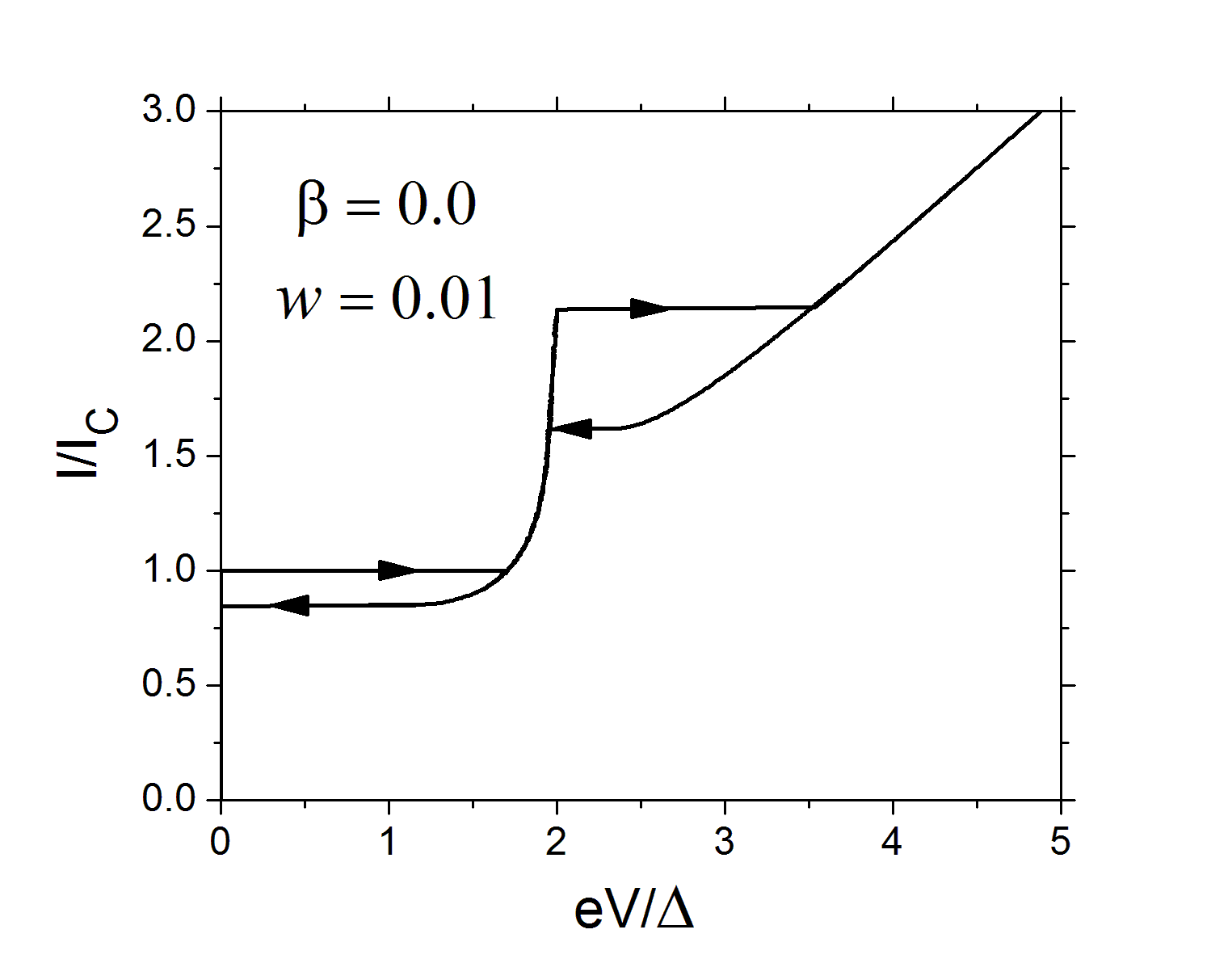}
\caption{{\protect\footnotesize {Current-voltage characteristic of a current-biased Josephson tunnel junction with negligible junction capacitance and equal energy gaps of the electrodes. } }}
\label{fig2} \end{figure}

I now discuss the results for the dc current-voltage characteristics (``$IV$ curves'') of a current-biased Josephson tunnel junction obtained by the direct numerical solution of the integro-differential equations (\ref{dim}) with the current kernels (\ref{dim2}) regularized according to Eq.~(\ref{gbr}). Although finite shunt resistance $R_S$ does not make the simulation of the junction dynamics any more difficult, to avoid extra parameters that are not relevant to our main purpose, only unshunted junctions are considered below. Figure \ref{fig1} shows an example of an $IV$ curve calculated for the typical value of the broadening parameter, $w=0.01$, and a moderate junction capacitance $\beta=0.3$. The main feature of the $IV$ curve is that it exhibits two hysteresis loops: one at voltages below the gap voltage $V_g=(\Delta_1+\Delta_2)/e$, and one -- above the gap. The hysteresis at $V<V_g$ is well-understood in terms of the inertia introduced into the phase dynamics by the junction electric capacitance $C$, and is viewed as the characteristic feature of the tunnel-junction behavior in the Josephson dynamics. Indeed, for large capacitance and weak dissipation characteristic for tunnel junctions, the bias current needs to be reduced well below the critical current in order to stop the time evolution of the phase. In contrast to the inertial hysteresis, to the best of author's knowledge, the above-the-gap hysteresis has not been analyzed or discussed previously, despite some indirect evidence in the literature \cite{mcdonald1976,zorin1983}.

\begin{figure}[t]
\centering
\includegraphics[width=0.48\textwidth]{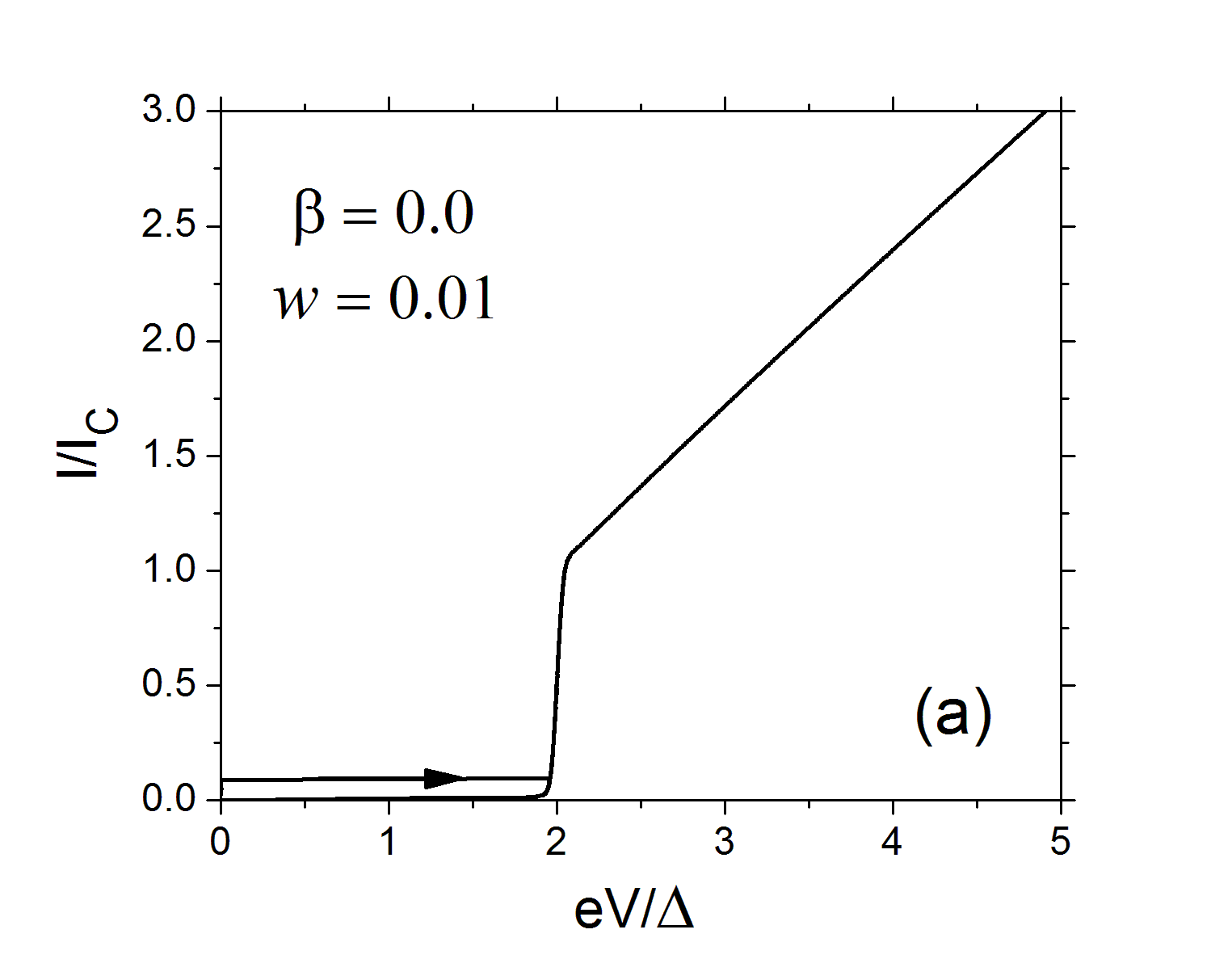} \\
\includegraphics[width=0.48\textwidth]{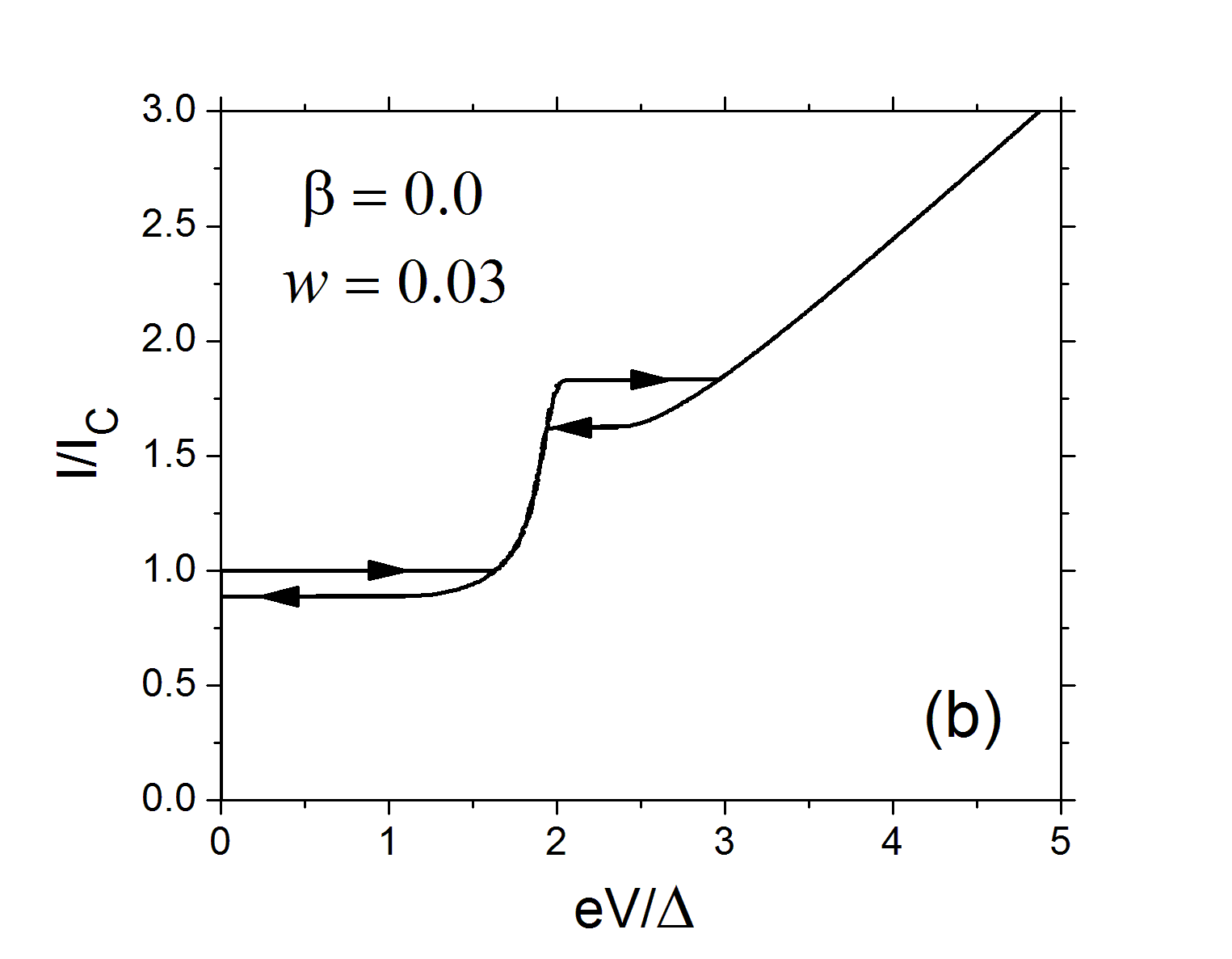}
\caption{{\protect\footnotesize {Current-voltage characteristics of a current-biased Josephson tunnel junction: (a) with suppressed supercurrent: $i_p(\tau) \rightarrow 0.1 i_p(\tau)$, and (b) with increased broadening of the gap singularity. } }}
\label{fig3} \end{figure}

Figure \ref{fig1} presents an example of the $IV$ curve of the ``asymmetric'' junction with different energy gaps, $\Delta_1=0.5\Delta_2$.  By making similar calculations for other gap ratios one can see that the difference of the energy gaps of the two junction electrodes does not have qualitative effects on the junction current-voltage characteristics. To simplify the discussion of the results, in what follows, I treat numerically only the case of equal gaps, $\Delta_1=\Delta_2\equiv \Delta$. Consistent with its inertial mechanism, the hysteresis at $V<V_g$ in the $IV$ curves becomes larger with increasing junction capacitance, i.e., increasing $\beta$. As can be seen from Fig.~\ref{fig2}, which shows the $IV$ curve for vanishing junction capacitance, $\beta=0$, the hysteresis at $V\gtrsim  V_g$ becomes, by contrast, more pronounced with decreasing capacitance $C$. With increasing $\beta$, this hysteresis shrinks, and one can see that for equal gaps and $w=0.01$, it disappears for $\beta>0.6$. Inertial hysteresis at $V<V_g$ is indeed smaller in Fig.~\ref{fig2}, but does not vanishes completely, as might have been expected for vanishing capacitance. The most probable reason for this is the reactive component of the tunnel current that results in a complex time evolution of the phase, which gives a non-vanishing effective correction to the junction capacitance.  To understand the origin of the hysteresis at $V\gtrsim V_g$, one can note from Fig.~\ref{fig2} that one, `` vertical'', side of the hysteresis is very close the gap voltage, $V\simeq V_g$. This suggests that the hysteresis is related to the Riedel singularity of the supercurrent in a tunnel junction. This consideration is supported by the $IV$ curves calculated for the junction dynamics with reduced pair component of the tunnel current (something that can be achieved in principle, e.g., with the help of a magnetic field) or increased broadening parameter $w$. As one can see from Fig.~\ref{fig3}, in both cases, above-the-gap hysteresis is indeed suppressed. Reduction of the pair component of the current leads also to a smaller inertial hysteresis because of the reduced junction critical current. All this makes the junction $IV$ curve shown in Fig.~\ref{fig3}a very close to just the dependence of the quasiparticle component of the tunnel current on voltage in the regime of the fixed dc bias voltage $V$. Extra (but still small) broadening of the gap singularity illustrated in Fig.~\ref{fig3}b does not affect the inertial hysteresis while reducing the hysteresis at $V\gtrsim V_g$. This confirms that the latter hysteresis has the same origin as the Riedel peak of the supercurrent at fixed bias voltage.

\begin{figure}[t]
\centering
\includegraphics[width=0.47\textwidth]{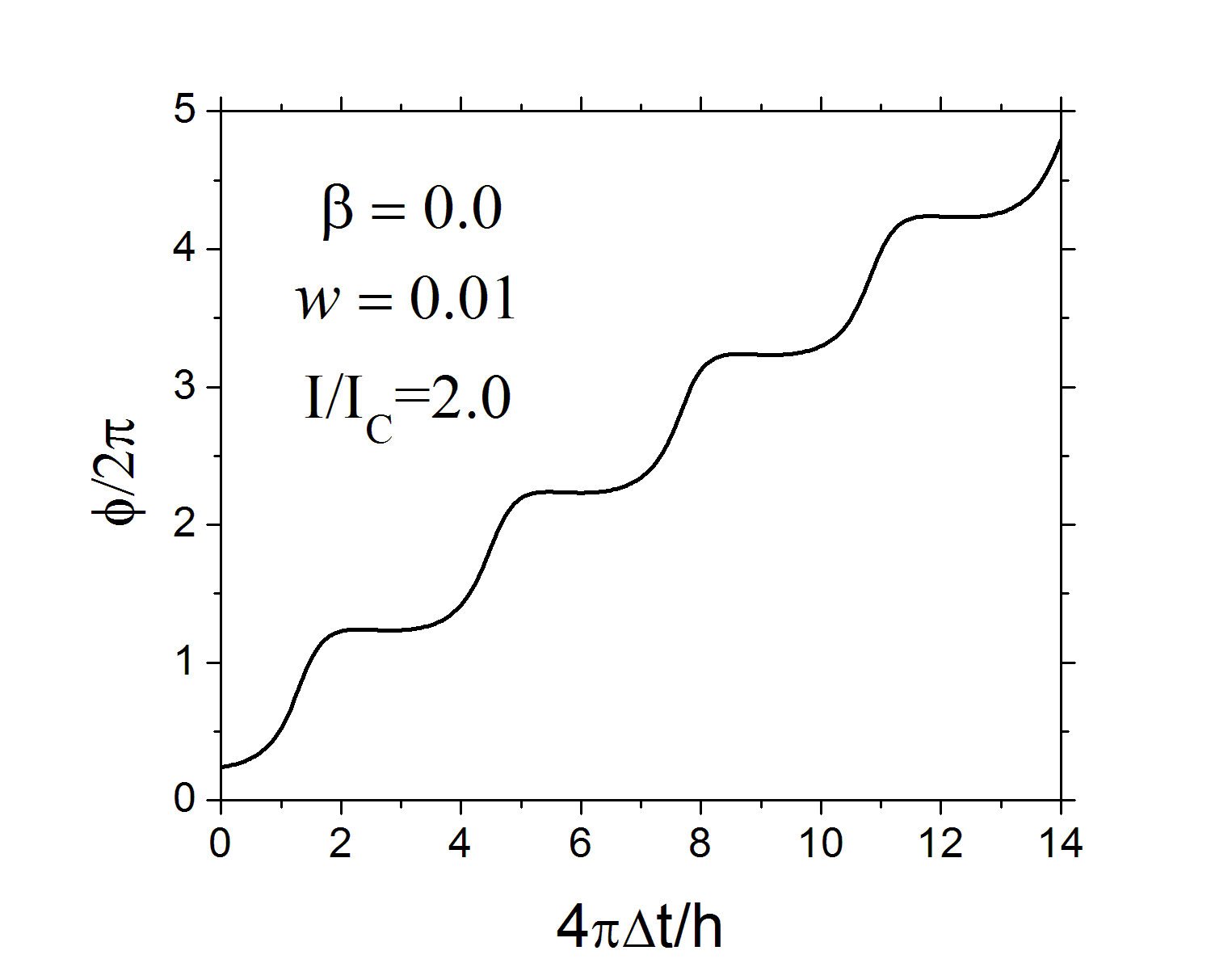}
\caption{{\protect\footnotesize {Typical time evolution of the Josephson phase across the current-biased Josephson tunnel junction on the vertical branch of the above-the-gap hysteresis. The plotted phase evolution corresponds to the point $I=2.0I_C$, $V\simeq 2\Delta/e$ on the $IV$ curve shown in Fig.~\ref{fig2}. Note that the time $t$ in this plot is normalized to the total gap frequency, i.e., the period of the gap oscillations is $2\pi$. The period of the Josephson oscillations, which have the frequency that is twice large than the gap frequency, is $\pi$, consistent with the plot. } }}
\label{fig4} \end{figure}

To develop a more detailed understanding of this hysteresis, it is helpful to look at the time evolution of the Josephson phase $\phi(t)$ on the vertical branch of the hysteresis, $V\simeq V_g$. An example of this evolution is shown in Fig.~\ref{fig4} which plots $\phi(t)$ for the junction with vanishing capacitance, $\beta=0$, biased on the vertical branch of the hysteresis: $I=2.0I_C$, $V\simeq 2\Delta/e$. One sees from this plot that the phase evolves in a step-like manner, with the junction spending large fraction of the Josephson oscillation period with $\phi(t)\simeq (\pi/2) \mbox{mod} (2\pi)$. Since $V\simeq V_g$ in this regime, the Josephson frequency $\omega_J=2eV/\hbar$ is simply related to the gap oscillations frequency, $\omega_J=2\Omega$, and as shown below, the integral of the asymptotics of the pair current indeed gives a non-vanishing contribution to the dc tunnel current.

In more details, we assume the simplest model of the phase evolution qualitatively consistent with the actual staircase-like time dependence shown in Fig.~\ref{fig4}:
\begin{equation}
\phi (t)=\frac{\pi}{2} + 2\pi n(t),
\label{e40} \end{equation}
where $n(t)$ is an integer that increases by 1 with each period $2\pi/\omega_J$ of the Josephson oscillations. For such time dependence of the phase, $\sin \phi(t) \equiv 1$, while $\sin [\phi (t)-\phi (t')]/2 \equiv 0$. This means that besides the Ohmic component, only the pair current gives nonvanishing contribution to the total tunnel current, and
\[ \sin \frac{\phi (t)+\phi(t')}{2} = \sin \phi(t) \cos \frac{\phi (t)-\phi (t')}{2}\]

\vspace*{-4ex}

\[ = \cos \pi [n(t)+n(t')]= (-1)^{n(t)+n(t')}\,  .\]

Under the assumption that the Riedel singularity is smeared only weakly, the pair component of the tunnel current is dominated in time domain by its large-time asymptotics (\ref{lt}). In this approximation, expression for the pair current in dimensional form is:
\begin{equation}
I(t)=\frac{\sqrt{\Delta_1\Delta_2}}{e R_N}\int \frac{d\eta}{\eta} \cos \Omega \eta \sin\frac{\phi(t)+\phi(t-\eta)
}{2}\, ,
\label{e41} \end{equation}
where the lower limit of the integral is set by condition of validity of the long-time asymptotics of the Bessel functions to $\eta \simeq \Omega^{-1}$, while the upper limit is determined by the width of the Riedel singularity, e.g., by Eq.~(\ref{gbr}): $\eta \simeq (w\Omega)^{-1}$. For small width of the Riedel singularity, $w\ll 1$, the integral in Eq.~(\ref{e41}) can be found in two steps. First, one averages out the rapidly-varying terms of frequency $\Omega$ in Eq.~(\ref{e41}) by integrating over one period of the gap oscillations neglecting the slow-varying factors, including the factor $1/\eta$. Then, one integrates the slow-varying terms with frequencies on the order of $w\Omega$, with high-frequency terms already averaged out.

On the vertical branch of the above-the-gap hysteresis, the deviation $\delta V$ of the dc voltage $V$ across the junction from the gap value is small. As will be seen soon, it is proportional to the width of the Riedel singularity, $\delta V\equiv V-V_g \propto w$. This means that the deviation of the Josephson frequency from $2\Omega$ is also small, and can be neglected when evaluating the integral (\ref{e41}) for the pair current over one period of the gap oscillations:
\[ I'(t)=\frac{\sqrt{\Delta_1\Delta_2}}{eR_N}\, \frac{\Omega}{2\pi} \int_0^{2\pi/\Omega}  d\eta \cos \Omega \eta \, (-1)^{n(t)+n(t-\eta)}\]

\vspace*{-3ex}

\begin{equation}
=\frac{2}{\pi} \frac{\sqrt{\Delta_1\Delta_2}}{eR_N} |\sin \Omega t|\; .
\label{e42} \end{equation}
In this expression, it is assumed that the time $t$ is counted starting at one of the jumps of the staircase-like evolution of the Josephson phase (Fig.~\ref{fig4}). To get the full magnitude of the pair current for the phase (\ref{e40}) we now need to take the second step and integrate the current (\ref{e42}) over the full range of the gap oscillations in Eq.~(\ref{e41}). In the approximation discussed above, the corresponding integral is:
\[ \int_{\Omega^{-1}}^{\infty} \frac{d\eta}{\eta} \cos [(\Omega-\omega_J/2)\eta]e^{-w\Omega \eta}=\ln\big[ w^2+
(\delta V/V_g)^2\big]^{-1/2} \, .\]
Combining these two results, we obtain an expression for the pair current on the vertical side of the above-the-gap hysteresis in the approximation (\ref{e40}) for $\phi(t)$:
\begin{equation}
I'(t)=\frac{2}{\pi} \frac{\sqrt{\Delta_1\Delta_2}}{eR_N} \ln\big[ w^2+ (\delta V/V_g)^2 \big]^{-1/2}
|\sin \Omega t|\; .
\label{e43} \end{equation}

Although, obviously, the assumption (\ref{e40}) of the sharp steps in the evolution of the phase $\phi(t)$ and the resulting time dependence (\ref{e43}) of the pair current do not provide an exact self-consistent solution of the evolution equations (\ref{dyn}), they describe the main qualitative feature of the smoother self-consistent evolution obtained by direct numerical solution of Eqs.~(\ref{dyn}) and shown in Fig.~\ref{fig4}. The staircase-like Josephson phase in Fig.~\ref{fig4} produces a finite dc component of the supercurrent in the junction, an estimate of which is obtained from Eq.~(\ref{e43}):
\begin{equation}
I=\frac{8}{\pi^2} I_R \, , \;\; I_R\equiv \frac{\sqrt{\Delta_1\Delta_2}}{2eR_N} \, \ln [ w^2+
(\delta V/V_g)^2]^{-1/2}  .
\label{e44} \end{equation}

By contrast, for fixed bias voltage $V$ across the junction, the regime that approximately describes the higher-voltage branch of the above-the gap hysteresis, the pair current does not have a dc component. If the phase evolves simply as $\phi(t)=\phi_0+\omega_J t$, the non-vanishing contribution to the pair current Eq.~(\ref{e41}) comes from the term:
\[ \sin \frac{\phi (t)+\phi(t')}{2} = \sin \phi(t) \cos \frac{\omega_J (t'-t)}{2} \, . \]
The same calculation as done above for the phase (\ref{e40}) shows then that the current (\ref{e41}) in this regime is:
\begin{equation}
I'(t)=I_R \sin \phi(t)\, ,
\label{e45} \end{equation}
and indeed averages to zero over time $t$. This also shows that the current $I_R$ introduced in Eq.~(\ref{e44}) has the meaning of the supercurrent amplitude in the junction voltage-biased in the vicinity of the Riedel singularity.

In summary, the difference between the vanishing and finite dc components of the pair current for the ``fixed-voltage'' linear evolution of the phase on one branch of the hysteresis, and the staircase-like evolution (Fig.~\ref{fig4}) on the other branch, explains the mechanism of the above-the-gap hysteresis in the current-bias Josephson tunnel junctions. This means that the hysteresis is closely related to the logarithmic gap singularity of the supercurrent in tunnel junctions. Since the magnitude of the dc component (\ref{e44}) of the supercurrent varies with the dc voltage $V$ across the junction on the scale $wV_g$, for small smearing $w$ of the gap singularity, the side of the hysteresis near the gap voltage looks nearly vertical (Fig.~\ref{fig2}), with the voltage $V$ pinned down to the gap voltage $V_g$. Also, from the point of view of this explanation, the fact that the hysteresis is suppressed with increasing junction capacitance is natural, since qualitatively, the capacitance makes the phase evolution more linear, suppressing the steps responsible for the dc part of the supercurrent.

\section{Conclusion}

In summary, a novel expression for the time-domain current kernels of a Josephson tunnel junctions between two ideal BCS
superconductors with in general different energy gaps is derived and used to simulate the dynamics of the current-biased junctions. The simulations make it possible to study in detail the current-voltage characteristics of such junctions and reveal their qualitatively new features, the main one -- an additional hysteresis at voltages above the gap voltage of the junction. The hysteresis is related to the Riedel singularity of the junction supercurrent and is explained in terms of the step-like time dependence of the Josephson oscillations at the gap voltage, that can be viewed as characteristic feature of the resonance between the Josephson oscillations and the gap oscillations with frequency $\Omega=(\Delta_1+\Delta_2)/\hbar$ in the pair current kernel of the junction, where $\Delta$'s are the energy gaps of the two junction electrodes. This regime of the gap resonance exists in the junctions with the relatively short $RC$ time constant, $\beta \equiv \Omega R_NC \lesssim 0.6$, where $R_N$ is the normal-state resistance, and $C$ -- capacitance of the junction. In typical Josephson tunnel junctions, $RC$ time constant is nearly independent of the junction area, and characterizes the ``critical current density'' $j$. Since $\beta$ is proportional to the gap frequency, the condition that $\beta$ is not too large is more difficult to fulfill in junctions employing superconductors with large energy gaps, like niobium that is used frequently to produce the large-scale circuits of superconductor electronics. Aluminum junctions, which are also quite standard, have significantly smaller gap $\Delta \simeq 0.2$ meV, so that the gap frequency is $\Omega \simeq 0.6$ THz and reaching the regime of the gap resonance, $\beta \lesssim 0.6$ requires junctions with the time constant $R_NC$ about 1 ps. High critical current density can be reached in tunnel junctions with insulator layer thickness below 1 nm. Combined with the relative dielectric constant $\epsilon \simeq 8$ of the standard aluminum oxide tunnel barrier, this gives the electric capacitance per unit area $C\simeq 7\cdot 10^{-14}$ F/$\mu$m$^2$, and $R_NC \simeq 1$ ps for the resistance of unit area $15$ Ohm$\cdot \mu$m$^2$. This translates into a critical current density $j\simeq 2$ kA/cm$^2$ that should be achievable in junctions which still preserve both the high-quality tunnel barriers and superconducting electrodes with sharp edges of the energy gap.

The author would like to thank Alexander Kirichenko and Pavel Shevchenko for sharing the unpublished results. This work was supported by the IARPA Supertools program through the Synopsys and Hypres subcontracts. The author would like to acknowledge Timur Fillipov, Deep Gupta, Amol Inamdar, Dmitri Kirichenko, Michail Kupriyuanov, Konstantin Likharev, Vasili Semenov, and Stephen Whiteley for the stimulating discussions of the results.

\appendix

\section{}

This appendix presents the derivation of the expressions for the current kernels discussed in Sec.~\ref{sec1}. Since the calculation of the tunnel current in a Josephson junction is described in details in the literature, only the steps leading specifically to Eqs.~(\ref{eq2}) are outlined.

Using the standard method of the tunnel Hamiltonian, the current $I_T(t)$ in a Josephson tunnel junction can be expressed as in Eq.~(\ref{eq1}) with the current kernels determined as follows. First, the standard evaluation of the pair creation/annihalation terms in the two junction electrodes (see, e.g., \cite{barone}) defines the pair-current kernel:
\begin{eqnarray}
I_p(t)=\frac{2\Delta_1 \Delta_2}{\pi e\hbar R_N} \mbox{Im} \big[F_1(t)F_2(t)\big]\, ,
\;\;\;\;\;\;\;\;\;\;\;\;\;\;\; \nonumber\\
F_j(t)=\! \int \frac{d\xi}{\xi} f(\xi) \rho_j(\xi) e^{-i\xi t/\hbar} \! , \rho_j(\xi)=\frac{|\xi| \theta(\xi^2-\Delta_j^2)}{\sqrt{\xi^2-\Delta_j^2}} . \;\;\;\;\;\;\;\;
\label{eq7} \end{eqnarray}
Here $\xi$ is the energy of the quasiparticle states of the BCS Hamiltonian, $f(\xi)$ is the Fermi distribution function at temperature $T$, and $\rho_j(\xi)$ is the superconducting density of states in the $j$th electrode. Taking into account in Eq.~(\ref{eq7}) that
\[\mbox{Im} \big[ e^{-i\xi t/\hbar}e^{-i\zeta t/\hbar}\big]=-\sin\frac{\xi t}{\hbar}
\cos\frac{\zeta t}{\hbar}-\cos\frac{\xi t}{\hbar}\sin\frac{\zeta t}{\hbar} \, ,\]
and
\[ f(\xi)+f(-\xi)=1\, ,\;\;\; f(\xi)-f(-\xi)=-\tanh(\xi/2T) \, ,\]
one directly obtains Eqs.~(\ref{eq2}) and (\ref{eq3}) for $I_p(t)$.

Similarly to Eq.~(\ref{eq7}) for the pair kernel, equation for the quasiparticle current kernel $I_{qp}(t)$ is:
\begin{eqnarray}
I_{qp}(t)=\frac{2}{\pi e\hbar R_N} \mbox{Im} \big[G_1(t)G_2(t)\big]\, ,
\nonumber \\ G_j(t)=\int d\xi f(\xi) \rho_j(\xi) e^{-i\xi t/\hbar} .
\label{eq8} \end{eqnarray}
The same transformations of Eq.~(\ref{eq8}) as for the Eq.~(\ref{eq7}) would give:
\begin{eqnarray}
I_{qp}(t)=\frac{2}{\pi e\hbar R_N} \int_0^{\infty} d\xi d\zeta \rho_1(\xi) \rho_2(\zeta) \;\;\;\;
\nonumber\\ \cdot \Big(\sin\frac{\xi t}{\hbar} \cos\frac{\zeta t}{\hbar} \tanh \frac{\xi}{2T} +\cos\frac{\xi t}{\hbar}\sin\frac{\zeta t}{\hbar}\tanh \frac{\zeta}{2T}\Big) , \;\;\;\;\;
\label{eq9} \end{eqnarray}
and reproduce directly the integral part of the quasiparticle kernel (\ref{eq2}) in terms of the function $J_1(z)$ and $A_1(z,\beta)$. This conclusion, however, misses the normal-state Ohmic contribution to the current determined by the singular part of the integrals for $J_1(z)$ and $A_1(z,\beta)$ which are not properly convergent. A more accurate transformation of the expression for $I_{qp}$ requires to explicitly separate the singular part of these integrals, something that can be achieved by subtracting the normal density of states [equal to 1 in the conventions of Eqs.~(\ref{eq7}) and (\ref{eq8})] from the superconducting ones:
\begin{equation}
\rho_1 \rho_2=(\rho_1-1)(\rho_2-1)+(\rho_1-1)+(\rho_2-1)+1
\label{eq10} \end{equation}
in Eq.~(\ref{eq8}).

Then, one needs to do the calculation similar to that for the pair current kernel (\ref{eq7}) individually for each term in the expansion (\ref{eq10}). For the last, normal-state term, repeating the steps similar to those in Eq.~(\ref{eq6}), one obtains:
\begin{equation}
G_N(t)\equiv \int d\xi f(\xi) e^{-i\xi t/\hbar}= \frac{i\pi T}{\sinh[(\pi Tt/\hbar)+i0]} \, . \label{eq11} \end{equation}
With this result, Eq.~(\ref{eq8}) shows that the normal-state contribution to the quasiparticle current kernel is:
\begin{eqnarray}
I_N (t)=\frac{2}{\pi e\hbar R_N} \mbox{Im} \big[G_N(t)\big]^2 \nonumber \\
=-\frac{2\pi T^2}{e\hbar R_N} \mbox{Im}\frac{1}{\sinh^2[(\pi Tt/\hbar)+i0]} \, ,
\nonumber \\ =\frac{2T}{eR_N} \frac{d}{dt}\mbox{Im} \coth [(\pi Tt/\hbar)+i0]
=-\frac{2\hbar}{e R_N} \delta'(t) . \;\;\;\;
\label{eq12} \end{eqnarray}
This expression can be transformed into the corresponding part of Eq.~(\ref{eq2}). To provide the context for this transformation, it should be mentioned that all current kernels discussed in this work are obtained by the perturbation theory in tunneling which leads to  Eq.~(\ref{eq1}) for the current. Although the structure of the perturbation theory itself ensures through Eq.~(\ref{eq1}) that the current response to the bias voltage has the causal structure, the perturbative current kernels do not have this property. As can be seen from the integral form of the current kernel components (\ref{bes}) and (\ref{eq3}) (not the final form in terms of the Bessel functions) the kernels $I_p (t)$, $I_{qp}(t)$ have the property
\[ I_p (-t)=-I_p (t)\, ,\;\;\;\;\;  I_{qp}(-t)=-I_{qp}(t) \, .  \]
They do not vanish for $t<0$ as would be natural for the causal response, and as indeed happens in the non-perturbative calculations (see, e.g., \cite{averin2020}).  For the regularized ``superconducting'' part of the current kernels this is a formal property, since only the $t>0$ parts enter expression for the current. The singular nature of the normal part (\ref{eq12}), however, makes it necessary to resolve the singularity using the relation
\[ 2\int^{\infty}_0 dt \delta(t)=1=\int^{\infty}_0 dt \delta(t-0)\, . \]
Equation (\ref{eq2}) is written in the form that avoids this discussion.

The next term in Eq.~(\ref{eq10}) one should deal with is $\rho-1$, i.e.
\begin{eqnarray}
\int d\xi f(\xi) [\rho(\xi)-1] e^{-i\xi t/\hbar}=\int_0^{\infty} d\xi [\rho(\xi)-1] (\cos\frac{\xi t}{\hbar} \nonumber \\
+i \sin\frac{\xi t}{\hbar} \tanh \frac{\xi}{2T}) \, .
\nonumber  \end{eqnarray}
Written in dimensionless form ($x=\xi/\Delta$, $z=\Delta t/\hbar$), its real part is:
\begin{equation}
\Delta \int_0^{\infty} dx\Big[\frac{x \theta(x-1)}{\sqrt{x^2-1}} -1\Big]\cos zx \, ,
\label{eq13} \end{equation}
and is absolutely convergent. The second term in the brackets in this expression affects the value of the integral only at $z=0$:
\[ \int_0^{\infty} dx\cos zxe^{-\epsilon x}\Big|_{\epsilon \rightarrow 0} =\pi \delta (z)\, ,\]
while the first term is the derivative of the integral (\ref{bes}) for $J_0(z)$. This gives us the following expression for $J_1(z)$ for $z>0$:
\[ J_1(z)=-\frac{2}{\pi} \int_1^{\infty} dx \frac{x \cos zx}{\sqrt{x^2-1}}\, .\]
For $z=0$, the value of $J_1(0)=0$ is determined by the full integral (\ref{eq13}) with removed divergence.

Making use of Eq.~(\ref{eq6}), one can express the dimensionless form of the imaginary part of the $\rho-1$ term as
\begin{equation}
\int_0^{\infty} dx\Big[\frac{x \theta(x-1)}{\sqrt{x^2-1}} -1\Big]\sin zx \tanh \beta x =-\frac{\pi}{2}(A_1+S)\, ,
\label{eq14} \end{equation}
where $A_1$ is defined in Eq.~(\ref{eq3}) and
\[ S=\frac{1}{\beta \sinh(\pi Tt/\hbar)}\, .\]
Since the total integral (\ref{eq14}) is absolutely convergent, we know that $(A_1+S)\Big|_{z \rightarrow 0}=0$, i.e. $A_1$ diverges in the same way as $S$ for small $t$.

Combining the results for the real and imaginary parts of the $\rho-1$ term we get:
\begin{equation}
\int d\xi f(\xi) [\rho(\xi)-1] e^{-i\xi t/\hbar} =-\frac{\pi \Delta}{2}[J_1(z)+i(A_1+S)]\, .
\label{eq15} \end{equation}
Also, Eq.~(\ref{eq11}) for $G_N(t)$ can be written as
\begin{equation}
G_N(t)= \frac{\pi \Delta}{2}\Big[\frac{2\hbar}{\Delta} \delta (t) +i S(t)\Big]\, .
\label{eq16} \end{equation}
Using Eqs.~(\ref{eq15}) and (\ref{eq16}) to evaluate the contributions to the quasiparticle current kernel (\ref{eq8}) of all three terms (besides the normal-state term) in the expansion (\ref{eq10}), one gets the following total:
\begin{eqnarray}
\frac{\pi^2\Delta_1\Delta_2}{4}\mbox{Im} \Big\{[J_1+ i(A_1+S)]\big|_{\Delta_1}[J_1+ i(A_1+S)]\big|_{\Delta_2} \nonumber \\
-[J_1+ i(A_1+S)]\big|_{\Delta_1}[\frac{2\hbar}{\Delta} \delta (t) +i S] \big|_{\Delta_2} \nonumber \\
-[J_1+ i(A_1+S)]\big|_{\Delta_2}[\frac{2\hbar}{\Delta_1} \delta (t) +i S]\big|_{\Delta_1} \Big\} \nonumber \end{eqnarray}
\vspace{-3ex}
\[=\frac{\pi^2\Delta_1\Delta_2}{4} \Big\{J_1\big(\frac{\Delta_1 t}{\hbar}\big) A_1\big(\frac{\Delta_2t}{ \hbar},\beta_2\big) + J_1 \big(\frac{\Delta_2 t}{\hbar}\big) A_1\big( \frac{\Delta_1 t}{\hbar},\beta_1\big) \Big\}, \]
where in the last line we took into account that $(A_1+S)=0$ for $t=0$ ensuring that the products $(A_1+S)\delta (t)$ vanish. Together with the normal-state contribution (\ref{eq12}) this equation gives the quasiparticle current kernel (\ref{eq2}).

\section{}

The main element of the numerical procedure used to simulate the Josephson dynamics of the current-biased tunnel junctions in Sec.~\ref{sec3} is the calculation of the Bessel functions that determine the current kernels (\ref{eq5}), and through them, the time-dependent current response of the junction. The four Bessel functions that enter the expressions (\ref{eq5}) for the current kernels were calculated from the standard expressions. For small values of the dimensionless argument $x$, one can use the power-series expansions (see, e.g., \cite{as})
\begin{equation}
J_n(x)= \sum_{m=0}^{\infty} \frac{(-1)^m}{m!(n+m)!}\Big(\frac{x}{2} \Big)^{2m+n}\, ,
\label{b1} \end{equation}
The current kernels (\ref{eq5}), include only the first two functions, $n=0,1$. In practice, expansion (\ref{b1}) gives converging results for $x \lesssim 30$. Keeping up to $m=100$ terms in the series one can obtain $J_{0,1}$ with accuracy better that $10^{-4}$ in this range of the argument, as one can check either by comparison to a different, asymptotic expansion of these functions discussed below, or, by comparison with the numerical tables in \cite{as} for the values of $x$ for which the tables are available.

\begin{figure}[t]
\centering
\includegraphics[width=0.46\textwidth]{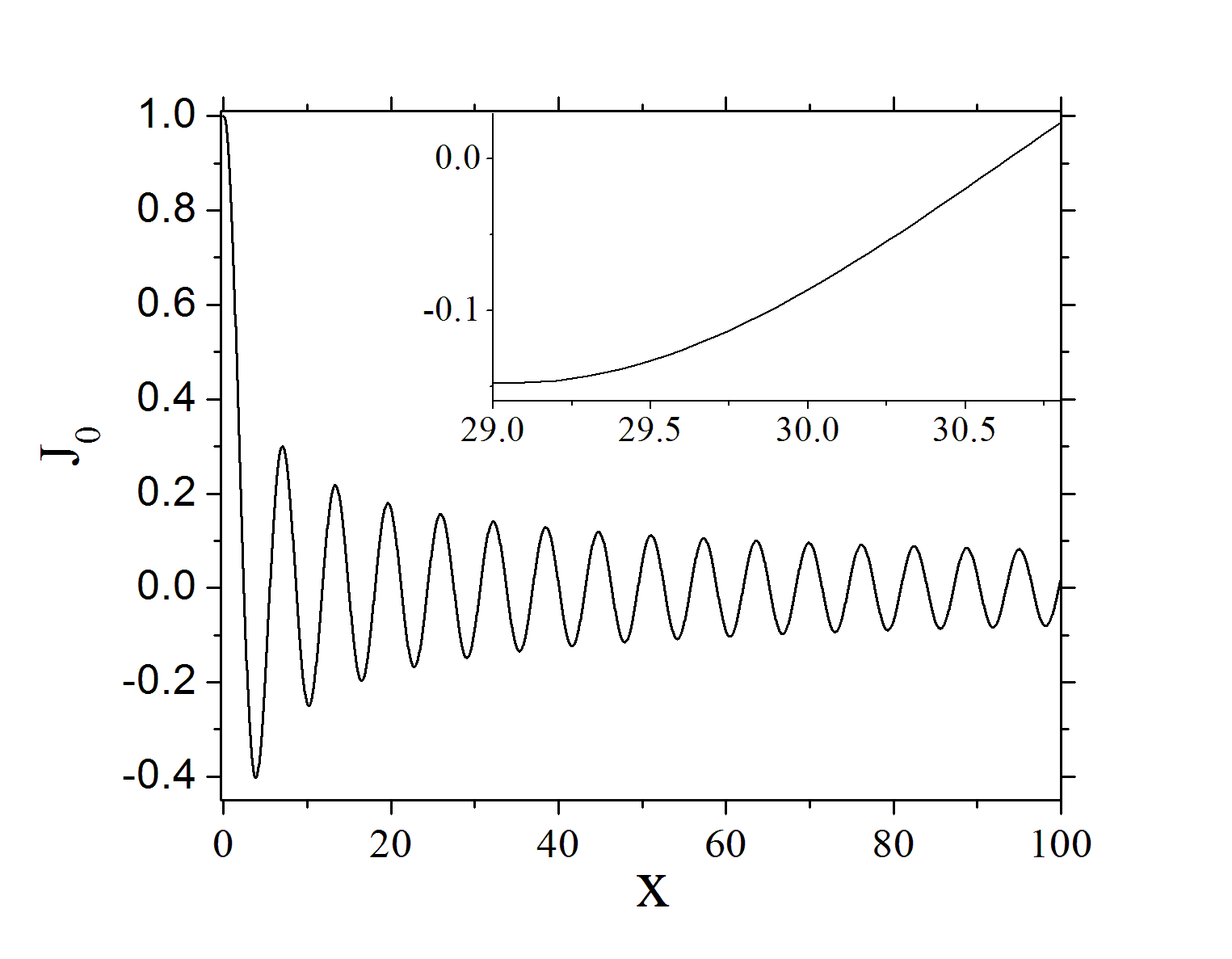} \\
\includegraphics[width=0.46\textwidth]{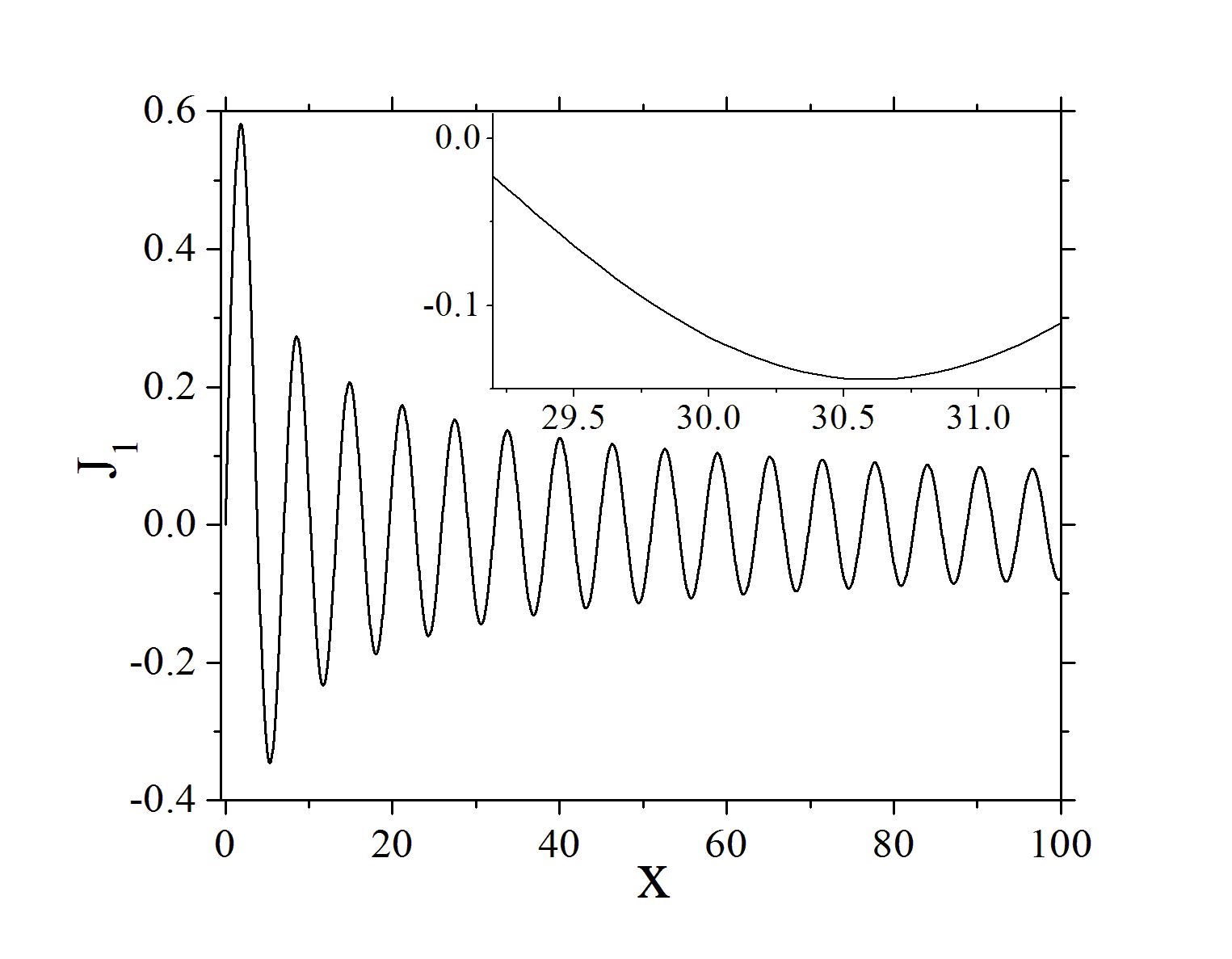}
\caption{{\protect\footnotesize {Two Bessel functions of the first kind calculated from Eqs.~(\ref{b1}) for $x\leq 30$ and from Eqs.~(\ref{b3}) and (\ref{b4}) for $x> 30$. The inserts show that even on a fine scale (on the order of $10^{-2}$), the two expansions, power series at small arguments and the asymptotic expansions at large arguments, agree very well, without a noticeable transition.  } }}
\label{fig5} \end{figure}

For the Bessel functions $Y_n(x)$ of the second kind, the power-series expansion similar to Eq.~(\ref{b1}) is
\begin{eqnarray}
Y_n(x)= \frac{2}{\pi} J_n(x)\Big(\ln \frac{x}{2}+\gamma \Big) -\frac{1}{\pi} \sum_{m=0}^{n-1} \frac{(n-1-m)!}{m!} \;\;\;\;\;\;\; \nonumber \\
\cdot  \Big(\frac{x}{2} \Big)^{2m-n} \!\!  -\frac{1}{\pi}  \sum_{m=0}^{\infty} \frac{(-1)^m (x/2)^{2m+n}}{m!(n+m)!}
\Big(\sum_{k=1}^{n+m} \! \frac{1}{k}+\sum_{k=1}^{m} \! \frac{1}{k} \Big) , \;\;\;\;\;\;\;\;
\label{b2} \end{eqnarray}
where $\gamma=0.5772156649$ is Euler's constant. The discussion of Eq.~(\ref{b1}) above applies to Eq.~(\ref{b2}) as well. The power series of up to $m=100$ terms converges for $x \lesssim 30$ and gives $Y_{0,1}$ with accuracy better that $10^{-4}$ in this range, as one can check by comparison to the asymptotic expansion of these functions.

\begin{figure}[t]
\centering
\includegraphics[width=0.46\textwidth]{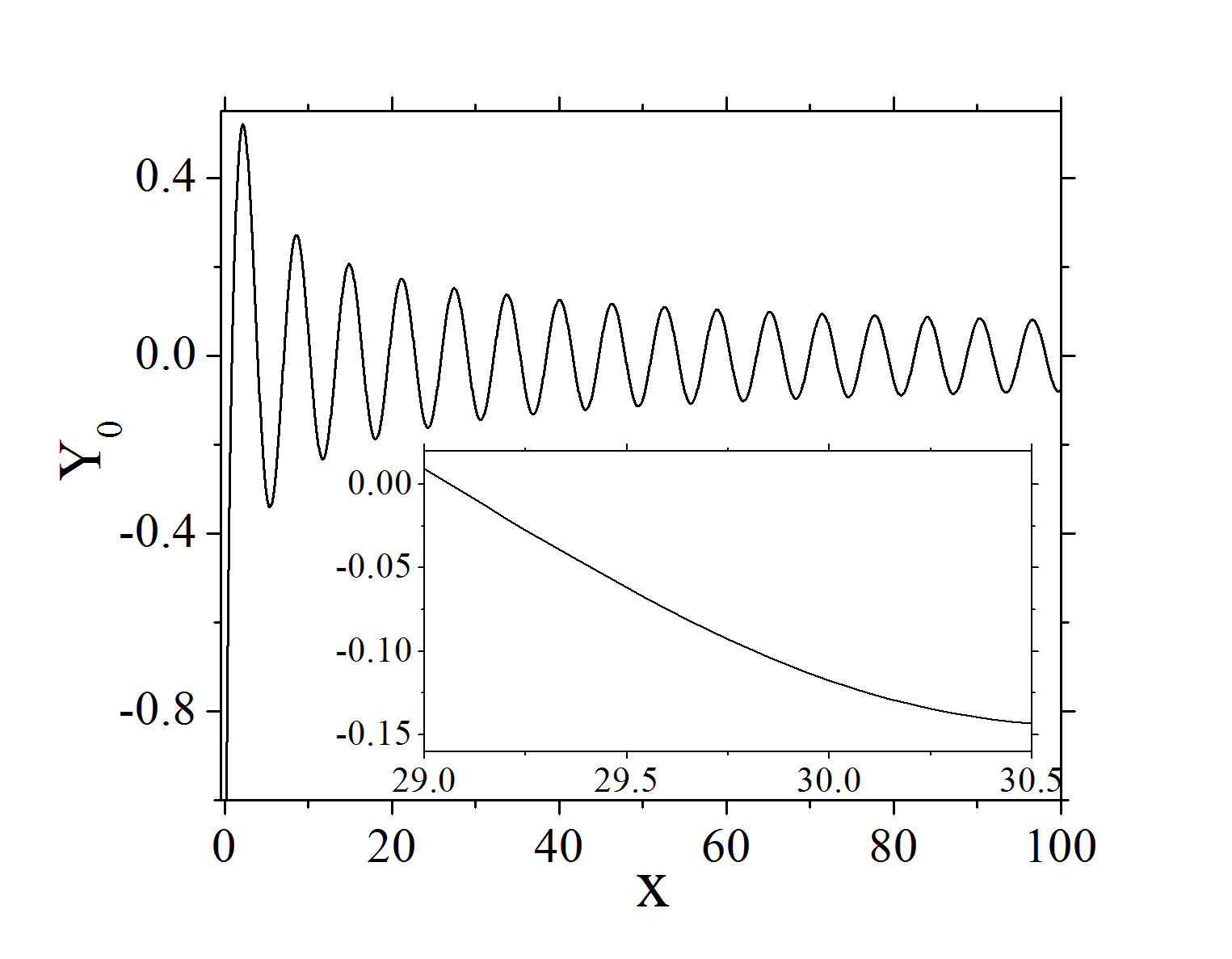} \\
\includegraphics[width=0.46\textwidth]{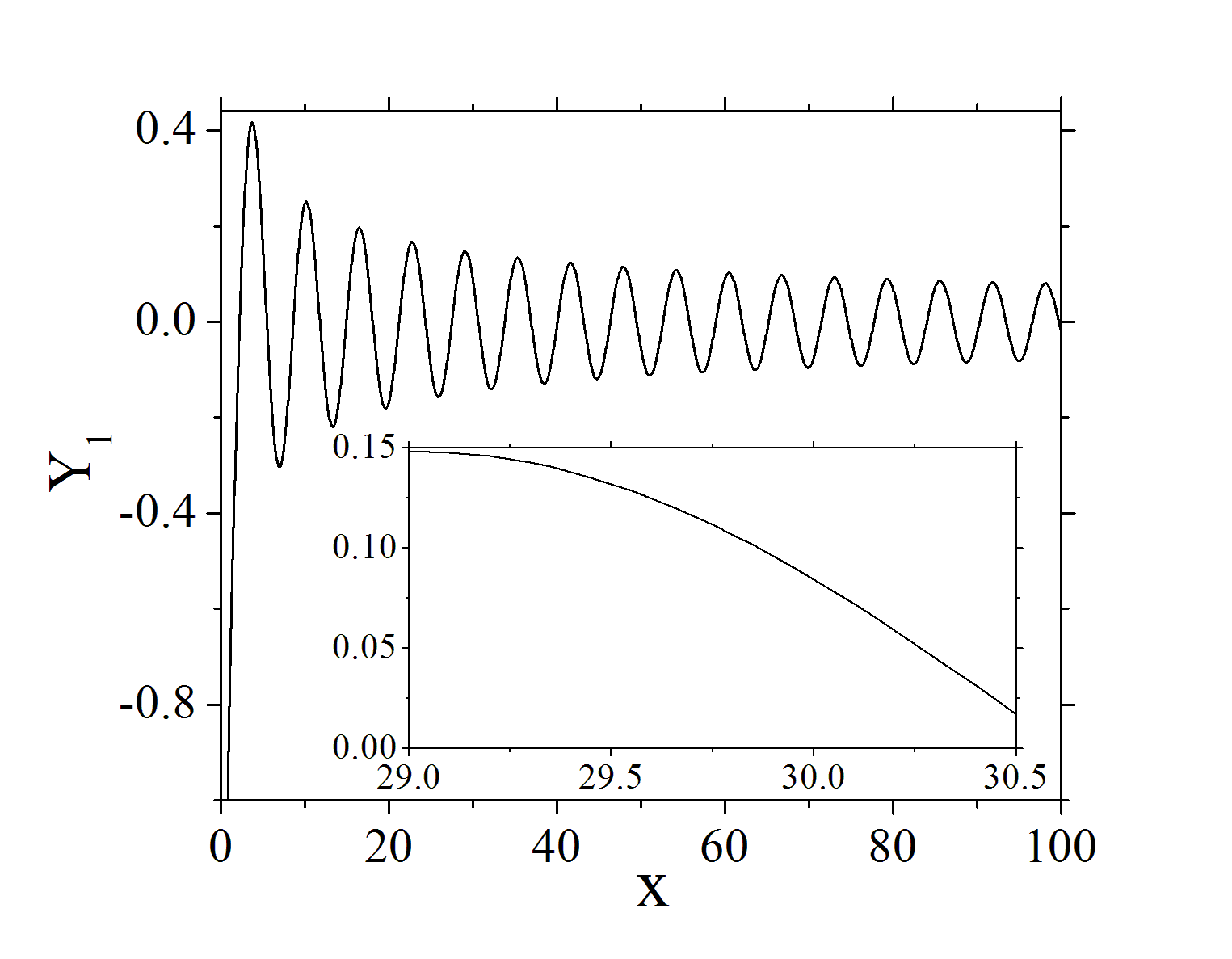}
\caption{{\protect\footnotesize {Two Bessel functions of the second kind calculated from Eqs.~(\ref{b2}) for $x\leq 30$ and from Eqs.~(\ref{b3}) and (\ref{b4}) for $x> 30$. Similarly to Fig.~\ref{fig5}, the inserts show that even on a fine scale on the order of $10^{-2}$, the two methods agree very well, without a noticeable transition.  } }}
\label{fig6} \end{figure}

Very slow convergence of the Bessel functions, which decay at large argument $x$ only as $x^{-1/2}$, implies that the integrals in Eq.~(\ref{eq1}) that determines the tunnel current do not converge by themselves under some conditions. For the realistic strength of the gap-smearing mechanisms which ensure convergence of these integrals, one needs to integrate the current kernels over the time range that extend to $x\sim 100$ in dimensionless units. This means that the power-series expansions for the Bessel functions discussed above are not sufficient, and one needs to use also the asymptotic expansions for large $x$. Moreover, to ensure that the asymptotic expansions match with sufficient accuracy the power-series expansions in the range $10 \lesssim x \lesssim 30$, where they are both applicable, one needs to go beyond the leading terms in the expansions. Keeping three terms should give the asymptotics with the accuracy on the order of $10^{-5}$ in the relevant range of the argument $x$. Then, for the four functions that enter the current kernels (\ref{eq5}), the appropriate asymptotic expansions are \cite{andrews}:
\begin{eqnarray}
J_0(x)= \Big(\frac{2}{\pi x}\Big)^{1/2} \big\{ P \cos \big(x-\frac{\pi}{4}\big) +Q \sin \big(x-\frac{\pi}{4}\big) \big\}
,\nonumber \\
Y_0(x)=  \Big(\frac{2}{\pi x}\Big)^{1/2} \big\{ P \sin \big(x-\frac{\pi}{4}\big) -Q \cos \big(x-\frac{\pi}{4}\big) \big\}
,\;\;\;\;\;\; \label{b3} \\
P=1-\frac{9}{128x^2}+\frac{3675}{2^{15} x^4}\, ,\;\;\;\; Q=\frac{1}{8x}-\frac{75}{1024 x^3} \, , \;\;\;\;\;\;\;\; \nonumber
\end{eqnarray}
and
\begin{eqnarray}
J_1(x)= \Big(\frac{2}{\pi x}\Big)^{1/2} \big\{ P' \cos \big(x-\frac{3\pi}{4}\big) +Q' \sin \big(x-\frac{3\pi}{4}\big) \big\}
 ,  \;\;\;\;\;\; \nonumber \\
Y_1(x)=  \Big(\frac{2}{\pi x}\Big)^{1/2} \big\{ P' \sin \big(x-\frac{3\pi}{4}\big) -Q' \cos \big(x-\frac{3\pi}{4}\big) \big\}
,\;\;\;\;\;\;\; \label{b4} \\
P'=1+\frac{15}{128x^2}-\frac{4725}{2^{15} x^4}\, ,\;\;\;\; Q'=-\frac{3}{8x}+\frac{105}{1024 x^3} \, . \;\;\;\;\;\;\;\; \nonumber
\end{eqnarray}

Four Bessel functions that determine the current response (\ref{eq1}) of a Josephson tunnel junction between the two ideal BCS superconductor with in general different energy gaps at zero temperature are shown in Figs.~\ref{fig5} and \ref{fig6}. The functions are calculated from the power series (\ref{b1})and (\ref{b2}) for small arguments, $x\leq 30$, and from the asymptotic expansions (\ref{b3}) and (\ref{b4}) for large arguments, $x> 30$. The main point illustrated by these plots, explicitly by the insets, is the agreement between the small-argument and large-argument expansions with very high accuracy. This accuracy is better than $10^{-4}$, as can roughly be estimated from these plots.

With the Bessel functions calculated accurately and in the arbitrary large range of the arguments as described above, Eqs.~(\ref{dim}) and  (\ref{dim2}) that govern the Josephson dynamics of the current-biased junctions can be solved directly by the simplest numerical procedures, integration of the current kernels at each time step to find the tunnel current, and then propagation of the values of the Josephson phase and the voltage actoss the junction according to the evolution equations. The results of these simulations are presented in Sec.~\ref{sec3}.

\end{document}